\documentclass{iau} 
\usepackage{graphicx}
\newcommand{\msun}{{\rm M}_\odot}

\newcommand{\gaea}{\sc{gaea}}
\title[Variable IMF in GAEA] {Variations of the stellar Initial Mass
  Function in Semi-Analytic Models: implications for the mass
  assembly of galaxies in the {\gaea} model.}

\author[Fabio Fontanot] {Fabio Fontanot$^1$}

\affiliation{$^1$ INAF - Astronomical Observatory of Trieste, via
  G.B. Tiepolo 11, I-34143 Trieste, Italy \\ email: {\tt
    fabio.fontanot@inaf.it}}

\pubyear{2018}
\volume{341}  %% insert here IAU Symposium No.
\jname{Challenges in Panchromatic Galaxy Modelling \\ with Next Generation Facilities}
\editors{M. Boquien, E. Lusso, C. Gruppioni, and P. Tissera, eds.}
\begin{document}

\maketitle

\begin{abstract}
A wealth of observations recently challenged the notion of a universal
stellar initial mass function (IMF) by showing evidences in favour of
a variability of this statistical indicator as a function of galaxy
properties. I present predictions from the semi-analytic model {\gaea}
(GAlaxy Evolution and Assembly), which features (a) a detailed
treatment of chemical enrichment, (b) an improved stellar feedback
scheme, and (c) implements theoretical prescriptions for IMF
variations. Our variable IMF realizations predict intrinsic stellar
masses and mass-to-light ratios larger than those estimated from
synthetic photometry assuming a universal IMF. This provides a
self-consistent interpretation for the observed mismatch between
photometrically inferred stellar masses of local early-type galaxies
and those derived by dynamical and spectroscopic studies. At higher
redshifts, the assumption of a variable IMF has a deep impact on our
ability to reconstruct the evolution of the galaxy stellar mass
function and the star formation history of galaxies.
\end{abstract}

\firstsection % if your document starts with a section,
\section{Introduction}

The IMF quantifies the number of stars formed per stellar mass bin in
a given episode of star formation. As such, it represents a key
parameter to characterise galaxy evolution from photometric
surveys. The IMF has long been assumed to have a universal shape,
mainly because of near-invariant results obtained for the closest star
forming regions in our Galaxy (see e.g. \cite{Kroupa01}). However,
recent observational developments put this notion in question. These
evidences came mainly trough two separated channels. A systematic
excess of the mass-to-light ratios derived using integral field
stellar kinematics with respect to photometric values based on a
universal IMF has been reported by many authors (see
e.g. \cite{Cappellari12}). Moreover, high-resolution spectroscopy of
spectral features sensitive to IMF variations (due of their dependence
on stellar surface gravity and/or effective temperature) suggests an
excess of low- relative to intermediate-mass stars, compared to the
universal IMF, in massive early-type galaxies (ETGs, see
e.g. \cite{Conroy13} and \cite{LaBarbera13}). In both cases, the
reported excess increases with increasing galaxy stellar mass
($M_\star$) and/or velocity dispersion. It is worth stressing that the
spectroscopic probe clearly favours a {\it bottom-heavy} scenario
(i.e. an excess of low-mass stars with respect to the universal IMF),
but it samples only to the low-mass end of the IMF ($<$1
$M_\odot$). On the other hand, the dynamical probe is sensible to the
whole shape of the IMF, but cannot distinguish between a {\it
  top-heavy} (i.e. an excess of high-mass stars with respect to the
universal IMF) or a {\it bottom-heavy} scenario. The observational
picture is further complicated by the radial IMF gradients reported
for local ETGs using integral field unit spectroscopy (see
e.g. \cite{Sarzi18}). These findings suggest that the low-mass stars
excess increases in the innermost regions of these galaxies. However,
these trends stand in apparent contradiction with the well-known
metallicity gradients in ETGs: the increasing $\alpha$-enhancement
towards the centre seems to require a larger contribution from Type-II
SNe, thus a larger fraction of high-mass stars.

\section{Variable IMF implementations in {\gaea}}

In my contribution, I present recent results obtained in the framework
of the semi-analytic model {\gaea} (GAlaxy Evolution and Assembly;
\cite{Hirschmann16}). This state-of-the-art model features key
improvements in the treatment of (a) chemical enrichment (i.e. it
follows the differential evolution of AGB stars, Type-Ia and Type II
SNe) and (b) stellar feedback (i.e. it implements an ejective feedback
prescription that combines results of high-resolution hydrodynamical
simulations with energy conservation arguments). This model is able to
provide a self-consistent picture of galaxy evolution, reproducing the
evolution of the galaxy stellar mass function (GSMF) and cosmic star
formation rate (SFR) up to $z\sim8-10$ \cite{Fontanot17b}.
\begin{figure}[h]
\begin{center}
 \includegraphics[width=15cm]{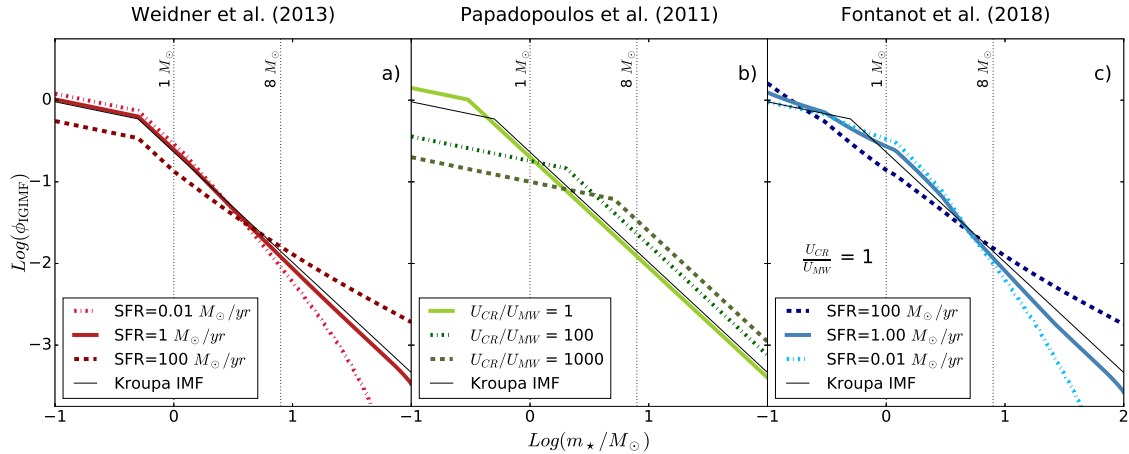} 
 \caption{Variable IMF implementations discussed in this talk (in all
   panels the thin solid line represents the universal IMF in the solar
   neighbourhood as in \cite{Kroupa01}). {\it Panel (a):} evolution of
   the IMF shape as a function of SFR, according to the IGIMF theory
   \cite{Weidner13a}.  {\it Panel (b):} evolution of the IMF shape as
   a function of cosmic ray energy densities, using the numerical
   results from \cite{Papadopoulos11}.  {\it Panel (c):} composite IMF
   evolution as a function of SFR and cosmic ray energy densities
   (only the $U_{\rm CR}=U_{\rm MW}$ case is shown) as proposed in
   \cite{Fontanot18b}. In all panels, each IMF is normalized to 1
   $\msun$ in the stellar mass interval 0.1-100 $\msun$.}
   \label{fig1}
\end{center}
\end{figure}

In a series of recent papers, we include in {\gaea} prescriptions that
link the variability of the IMF to physical properties of model
galaxies. In particular, in \cite{Fontanot17a} we test the integral
galaxy-wide IMF theory (IGIMF - \cite{Weidner13a} - Fig.~\ref{fig1},
{\it panel (a)}): this analytic model predicts an evolution of the
high-mass-end slope of the IMF as a function of the global SFR of
galaxies. In \cite{Fontanot18a}, we build on the results of numerical
simulations from \cite{Papadopoulos11} to relate the position of the
knee of the IMF with the cosmic ray density field ($U_{\rm CR}$,
normalized to the Milky Way reference value $U_{\rm MW}$) associated
with model galaxies (Fig.~\ref{fig1}, {\it panel (b)}). In a variable
IMF scenario, it is only possible to calibrate the models against
photometric data; in detail we consider $z=0$ luminosity functions
(LF) in the $g-$, $r-$, $i-$, and $K-$band, and the redshift evolution
of the $K$- and $V-$band LFs. In order to compute self-consistent
synthetic photometry for our variable IMF realizations, we construct
Single Stellar Populations libraries (for different ages and
metallicities) corresponding to each assumed IMF shape, using an
updated version of \cite{BC03} models. Synthetic photometry represents
the only observable we can directly compare with observational
constraints. Indeed, most of the estimates for physical properties of
galaxies available in the literature are derived under the hypothesis
of a universal IMF, and it could be problematic to compare them with
the {\it intrinsic} properties of model galaxies, as predicted by our
variable IMF runs. To better explain this issue, in the following we
focus on galaxy stellar masses. We define a photometrically-equivalent
{\it apparent} stellar mass ($M_\star^{\rm app}$) by applying to our
synthetic photometry mass-to-light versus colour scalings typically
used in photometric surveys (see e.g. \cite{Zibetti09}). In this way,
we derive the $M_\star$ an observer would associate to our synthetic
photometry assuming a universal IMF.

\begin{figure}[t]
\begin{center}
 \includegraphics[width=12cm]{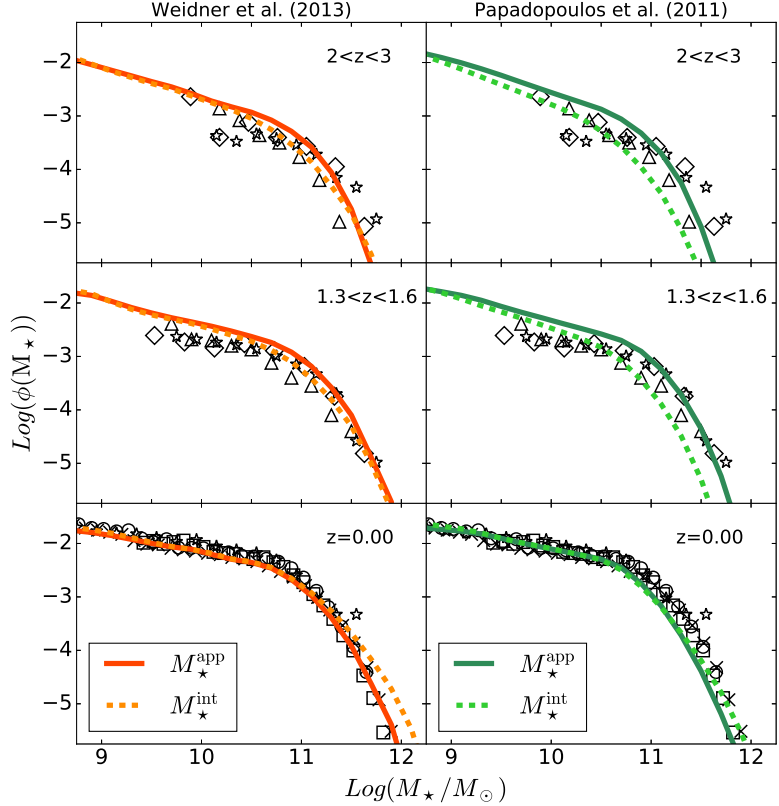} 
 \caption{Evolution of the GSMF. In each panel, solid lines correspond
   to the GSMF as a function of the apparent
   photometrically-equivalent stellar mass $M_\star^{\rm app}$, while
   dot-dashed lines show the GSMF as a function of the intrinsic galaxy
   stellar mass $M_\star^{\rm int}$. Dark points show a collection of
   observational measurements as in \cite{Fontanot09b}. }
   \label{fig2}
\end{center}
\end{figure}

\section{Results}

Both our variable IMF realizations provide consistent results in terms
of galaxy properties and their evolution. In particular, they are able
to naturally solve the long-standing problem of $\alpha$-enhancement
in massive ETG (see e.g. Fig.B1 in \cite[Fontanot et al.,
  2018]{Fontanot18a}). We compare the intrinsic stellar masses
$M_\star^{\rm int}$ as predicted by the models with $M_\star^{\rm
  app}$ and we find that our realisations are able to reproduce the
excess of dynamical mass (that closely correspond to $M_\star^{\rm
  int}$) with respect to the photometrically derived stellar mass (see
e.g. Fig.B3 in \cite[Fontanot et al., 2018]{Fontanot18a}). In the
framework of photometric surveys, the most interesting results involve
the study of galaxy evolution, as traced by statistical estimators
such as the GSMF. In this case, the effect of assuming a variable IMF
might be dramatic, as shown in Fig.~\ref{fig2}. Solid lines represent
the GSMF evolution derived using $M_\star^{\rm app}$: in most cases it
agrees quite well with the observational estimates (also obtained from
photometric surveys assuming a universal IMF). On the other hand,
dashed lines correspond to the intrinsic evolution of the GSMF in the
same {\gaea} runs: it is clear from this comparison that the variable
IMF hypothesis can change our understanding of galaxy evolution, as
seen using photometric surveys. In particular, the real GSMF might
differ significantly from estimates based on photometry, assuming a
universal IMF.

It is worth stressing that all variable IMF approaches we consider so
far do not predict a change in the low-mass end of the IMF. Therefore,
they wouldn't be able to reproduce the spectral features observed in
ETG spectra. Nonetheless, these models are able to reproduce the metal
enrichment of ETGs better than universal IMF runs.  In
\cite{Fontanot18b} we present a new derivation for the variable IMF,
that combines the IGIMF theory with the numerical results of
\cite{Papadopoulos11}. We assume the IMF shape depends on both the SFR
and $U_{\rm CR}$.  As a representative example, in Fig.~\ref{fig1},
{\it panel (c)}, we show the evolution of the IMF shape as function of
SFR, for a MW-like $U_{\rm CR}$. This scenario provides a possible
explanation for the observed IMF and metal gradients in ETGs. Indeed,
at increasing SFR the predicted IMF shape becomes {\it the same time}
shallower at the high-mass end (which implies a larger number of
massive stars) and steeper at the low-mass end (thus predicting an
excess of low-mass stars).

\end{document}